\documentclass[aps,prl,twocolumn,floatfix]{revtex4}

\usepackage{graphicx}
\usepackage{dcolumn}
\usepackage{bm}

\newcommand{\pdiffl}[2]{\frac{\partial #1}{\partial #2}}

\begin{document}


\title{Shock Hugoniot of diamond from 3 to 80\,TPa}

\date{August 14, 2020; revisions to March 29, 2022
  -- LLNL-JRNL-815773}

\author{Damian C. Swift}
\email{dswift@llnl.gov}
\affiliation{%
   Lawrence Livermore National Laboratory,
   7000 East Avenue, Livermore, California 94550, USA
}
\author{Andrea L. Kritcher}
\affiliation{%
   Lawrence Livermore National Laboratory,
   7000 East Avenue, Livermore, California 94550, USA
}
\author{Amy Lazicki}
\affiliation{%
   Lawrence Livermore National Laboratory,
   7000 East Avenue, Livermore, California 94550, USA
}
\author{James A. Hawreliak\footnote{%
Current affiliation:
Washington State University
}}
\affiliation{%
   Lawrence Livermore National Laboratory,
   7000 East Avenue, Livermore, California 94550, USA
}
\author{Tilo D\"oppner}
\affiliation{%
   Lawrence Livermore National Laboratory,
   7000 East Avenue, Livermore, California 94550, USA
}
\author{Heather D. Whitley}
\affiliation{%
   Lawrence Livermore National Laboratory,
   7000 East Avenue, Livermore, California 94550, USA
}
\author{Joseph Nilsen}
\affiliation{%
   Lawrence Livermore National Laboratory,
   7000 East Avenue, Livermore, California 94550, USA
}
\author{Benjamin Bachmann}
\affiliation{%
   Lawrence Livermore National Laboratory,
   7000 East Avenue, Livermore, California 94550, USA
}
\author{Michael MacDonald}
\affiliation{%
   Lawrence Livermore National Laboratory,
   7000 East Avenue, Livermore, California 94550, USA
}
\author{Brian Maddox}
\affiliation{%
   Lawrence Livermore National Laboratory,
   7000 East Avenue, Livermore, California 94550, USA
}
\author{Natalie Kostinski}
\affiliation{%
   Lawrence Livermore National Laboratory,
   7000 East Avenue, Livermore, California 94550, USA
}
\author{Siegfried Glenzer}
\affiliation{%
   SLAC National Accelerator Laboratory,
   Menlo Park, California 94025, USA
}
\author{Stephen D. Rothman}
\affiliation{%
   Atomic Weapons Establishment,
   Aldermaston, Berkshire, RG7~4PR, UK
}
\author{Dominik Kraus\footnote{%
Current affiliation: Helmholtz Zentrum, Dresden, Germany
}}
\affiliation{%
   University of California -- Berkeley,
   California 94720, USA
}
\author{Gilbert W. Collins\footnote{%
Current affiliation:
University of Rochester
}}
\affiliation{%
   Lawrence Livermore National Laboratory,
   7000 East Avenue, Livermore, California 94550, USA
}
\author{Roger W. Falcone}
\affiliation{%
   University of California -- Berkeley,
   California 94720, USA
}

\begin{abstract}
The principal Hugoniot of carbon, initially diamond,
was measured from 3 to 80\,TPa (30 to 800 million atmospheres),
the highest pressure ever achieved,
using radiography of spherically-converging shocks.
The shocks were generated by ablation of a plastic coating by soft x-rays
in a laser-heated hohlraum at the National Ignition Facility (NIF).
Experiments were performed with low and high drive powers, spanning different
but overlapping pressure ranges.
The radius-time history of the shock, and the profile of mass density behind,
were determined by profile-matching from a time-resolved x-ray radiograph
across the diameter of the sphere.
Above $\sim$50\,TPa, the heating induced by the shock was great enough to ionize
a significant fraction of $K$-shell electrons, reducing the opacity to
the 10.2\,keV probe x-rays.
The opacity and mass density were deduced simultaneously using the constraint
that the total mass of the sample was constant.
The Hugoniot and opacity were consistent with density functional theory calculations
of the electronic states and equation of state (EOS), and varied significantly
from theoretical Hugoniots based on Thomas-Fermi theory.
Theoretical models used to predict the compressibility of diamond ablator experiments at the NIF,
producing the highest neutron yields so far from inertial confinement fusion experiments,
are qualitatively consistent with our EOS measurements
but appear to overpredict the compressibility slightly.
These measurements help to evaluate theoretical techniques and
constrain wide-range EOS models applicable to white dwarf stars,
which are the ultimate evolutionary form of at least 97\%\ of stars in the
galaxy.
\end{abstract}


\maketitle

\section{Introduction}
Carbon (C) is widespread in nature and thought to be 
the fourth most common element in the universe \cite{AllendePrieto2002}.
Formed from the fusion of He in giant stars and at the end of the red giant
phase of stars like the sun, C accumulates at high levels in the core 
and is a major component of white dwarf (WD) stars, 
where it is predicted to segregate
and crystallize on cooling \cite{vanHorn1968},
or in the crust of accreting neutron stars \cite{Horowitz2007}.
Although there is little debate about its properties at pressures high enough 
for the electrons to become degenerate, they are less certain when atoms are
partially ionized. 
The ionization behavior affects the equation of state (EOS),
radiative opacities, and diffusion coefficients,
and is smoothed out in Thomas-Fermi (TF) models used in
predicting most EOS models for WDs,
which are the final state of the vast majority of stars in our galaxy
\cite{Fontaine2001}.
This uncertainty limits our understanding of the convection zone in WDs and
thus their cooling and evolution.

States of matter at elevated pressure and temperature are often generated
in shock wave experiments \cite{shock}, 
where the high-pressure matter is confined inertially
(i.e. by the finite time required for the compressed components
to disassemble) and so the pressures achieved are not limited by the 
strength of surrounding components as is the case with static presses
\cite{dac}.
Dynamic loading experiments are ubiquitous for studies of warm dense
matter with pressures in excess of 1\,TPa.

Large pulsed lasers such as the National Ignition Facility (NIF)
can be used to induce pressures in excess of 10\,TPa, which are
of interest for studies of massive exoplanets, brown dwarfs, and stars,
as well as engineering problems such as inertial confinement fusion (ICF).
We have previously reported the use of radiography of a spherically-converging
shock to deduce a range of states along the shock Hugoniot of a sample
\cite{Kritcher2014,Doeppner2018,Swift2018,Kritcher2020,Swift2021}, 
up to $\sim$40\,TPa in polystyrene,
and thus to constrain the EOS.
The x-ray source was a foil, laser-heated to a plasma emitting strongly in
the kilovolt regime.
In the results reported previously, the technique used to analyze the
radiograph used a model of the variation in mass density behind the shock
wave expressed in terms of radius and time.
For shock pressures high enough to affect the $K$-shell electrons and hence
the opacity to kilovolt-scale photons, a Lagrangian feature was
used to constrain the total mass of the sample in the density model,
and hence deduce the changing opacity at the shock front between instants
of time \cite{Kritcher2020,Swift2021}.

In the work reported here, we apply the same experimental technique to samples
of carbon (initially diamond),
deducing EOS data to significantly higher pressures than in polystyrene:
$\sim$80\,TPa, which is the highest pressure at which
material properties have been measured in the laboratory.
We use a modified analysis method, parameterizing the model
in terms of functions related to the EOS and reconstructing the density model
indirectly, rather than the other way around.
This method gives a smaller uncertainty and additional EOS data.
As before, data obtained from the experiment are absolute -- they are not made
relative to the properties of a reference material -- and the opacity was
deduced simultaneously with the mass density.
Absolute measurements of diamond are particularly impactful
as it is used widely as a reference or benchmark in measurements of 
other materials.

\section{Experimental configuration}
\begin{figure}
\begin{center}
\includegraphics[scale=0.30]{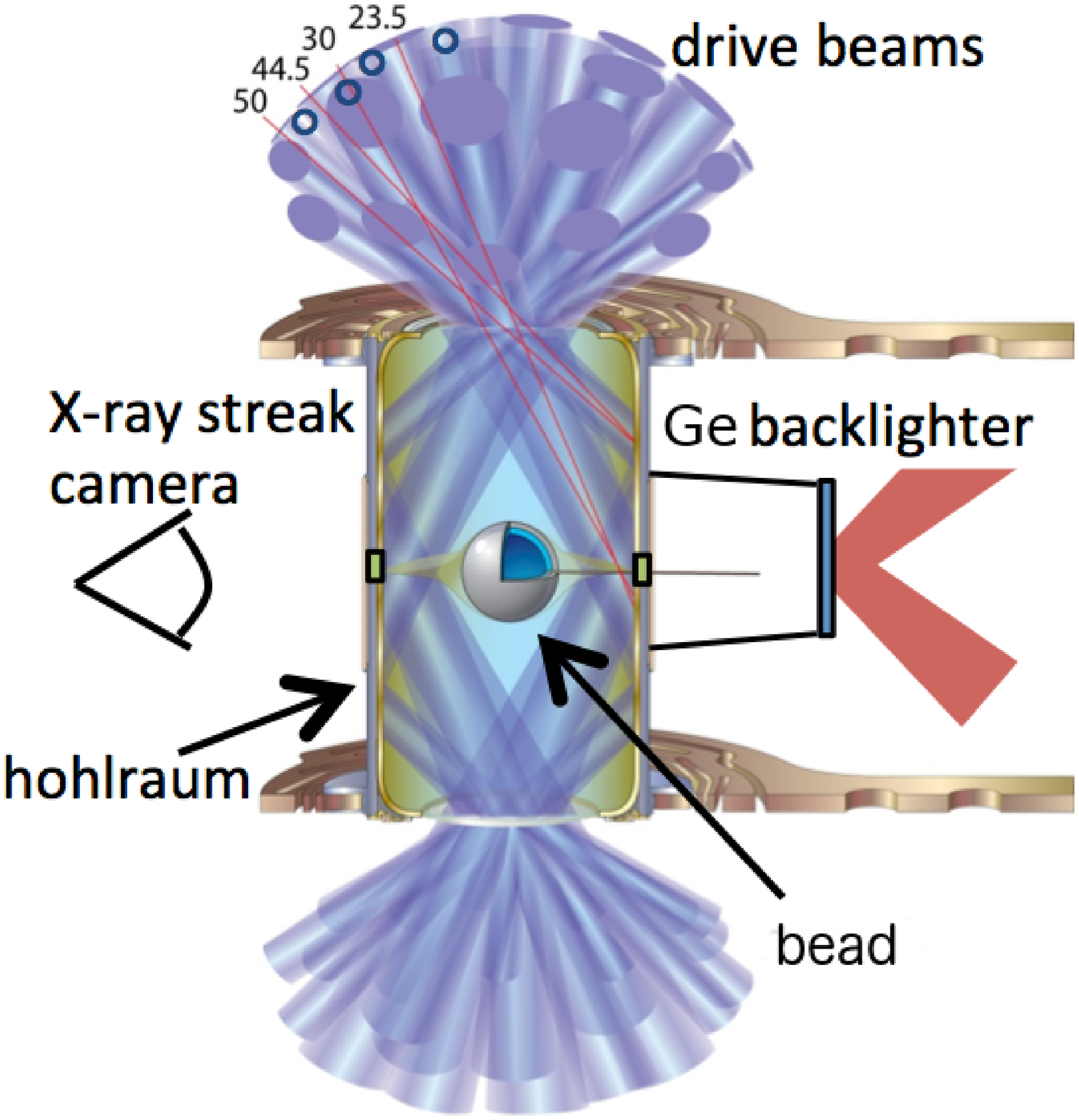}
\includegraphics[scale=0.80]{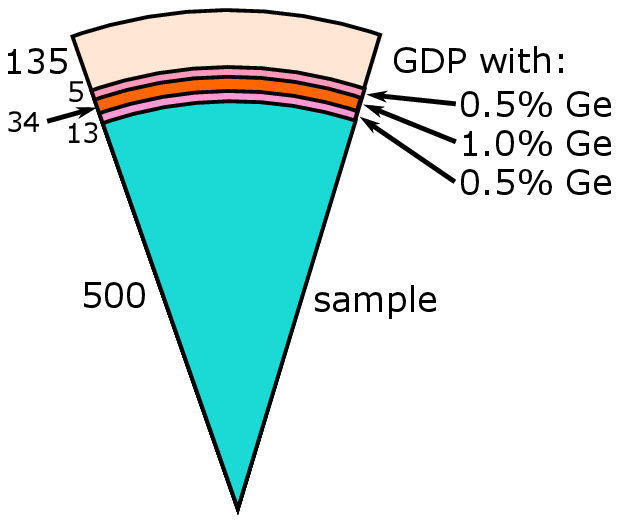}
\end{center}
\caption{Schematic of hohlraum-driven converging-shock experiment.
   Wedge diagram shows sequence of shells comprising spherical target bead,
   with radial thickness in microns.}
\label{fig:exptschem}
\end{figure}

The experimental configuration was as described previously
\cite{Kritcher2014,Doeppner2018,Kritcher2020}, and is summarized here for convenience.
Diamond spheres, 1000\,$\mu$m in diameter, were obtained from Dutch Diamond Technologies B.V.
A glow-discharge polymer (GDP) layer, 150\,$\mu$m thick, was deposited on the spheres
to act as an ablator.
The coated sphere was mounted within a Au hohlraum \cite{hohlraum},
30\,$\mu$m thick, 5.75\,mm diameter and 9.42\,mm high,
with a gas fill of 0.03\,mg/cm$^3$ He to impede filling of the hohlraum by ablated Au.
Up to 176 beams of the NIF laser were used to heat the hohlraum.
The resulting soft x-ray field within the hohlraum ablated the GDP,
driving a shock into the bead.
The overall configuration and laser pulses were based on ICF designs,
to take advantage of synergies in fabrication and also the large development
effort performed to give uniform drive conditions over the surface of the
bead \cite{icf}.
(Fig.~\ref{fig:exptschem}.)

We consider data from two experiments, N161016-3 and N140529-2, in which the
laser drive was 300 and 800\,kJ respectively.
The latter was based on the `high foot' ICF drive \cite{icf,Hopkins2015};
the former was the foot of the drive continued for 5\,ns.
The temperature history of soft x-rays in the hohlraum was calculated using
radiation hydrodynamics and measured by the {\sc dante} filtered diode system
\cite{Dewald2004}.
The peak temperature was around 205\,eV for the low drive and 275\,eV for the high.

\begin{figure}
\begin{center}\includegraphics[scale=0.65]{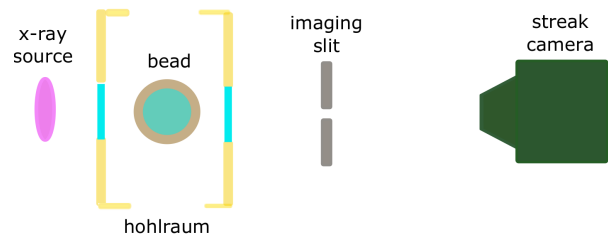}\end{center}
\caption{Radiograpic configuration (not to scale).}
\label{fig:rgconfig}
\end{figure}

The shock wave induced by ablation of the GDP strengthened as it propagated
toward the center of the sample.
X-ray radiography was used to measure the variation of attenuation
across the diameter of the bead, from which the shock trajectory,
mass distribution and opacity in the sample could be deduced as described
below.
The x-ray source was a Ge foil, heated by eight beams to produce a plasma
that emitted strong He-like radiation at 10.2\,kV, with duration $\sim$7\,ns.
Slits were cut in the hohlraum wall to enable transmission of the x-rays
through the sample; diamond windows were used to impede
slit closure by ablated Au.
The transmitted x-rays were imaged through a slit in a Ta foil onto 
an x-ray streak camera (Fig.~\ref{fig:rgconfig}).

The streak radiograph was used to reconstruct the radial distribution
of mass density, as a function of time.
As described previously \cite{Swift2018}, the presence of undisturbed material
ahead of the shock provided a strong constraint on the inference of
the change in attenuation across the shock front.
In order to take advantage of this constraint, the analysis was performed by
adjusting a parameterized representation of the distribution of mass density
until the corresponding simulated radiograph matched the measured radiograph.

In shot N140529-2, the shock became strong enough that the opacity of shocked
material to the probe x-rays decreased significantly.

\begin{figure}
\begin{center}\includegraphics[width=\columnwidth]{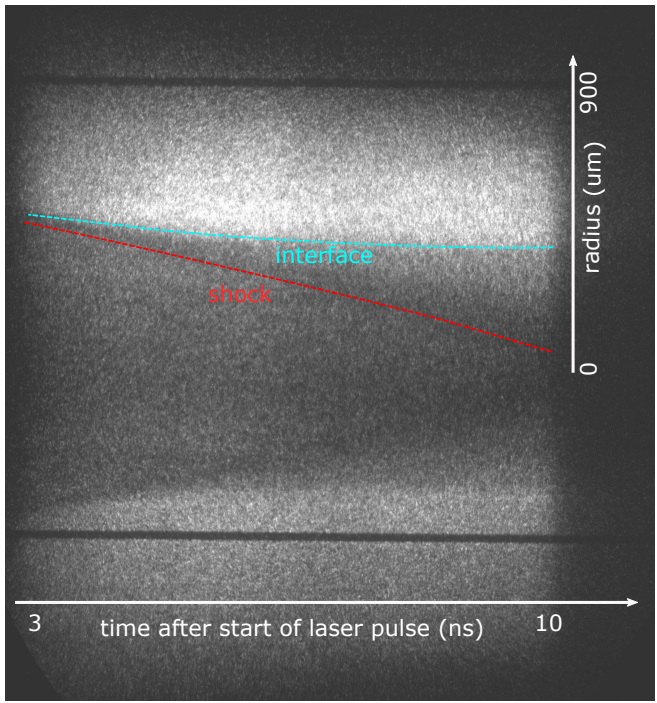}\end{center}
\caption{X-ray streak radiograph, NIF shot N161016-3 (low drive).}
\label{fig:rg161016}
\end{figure}

\begin{figure}
\begin{center}\includegraphics[width=\columnwidth]{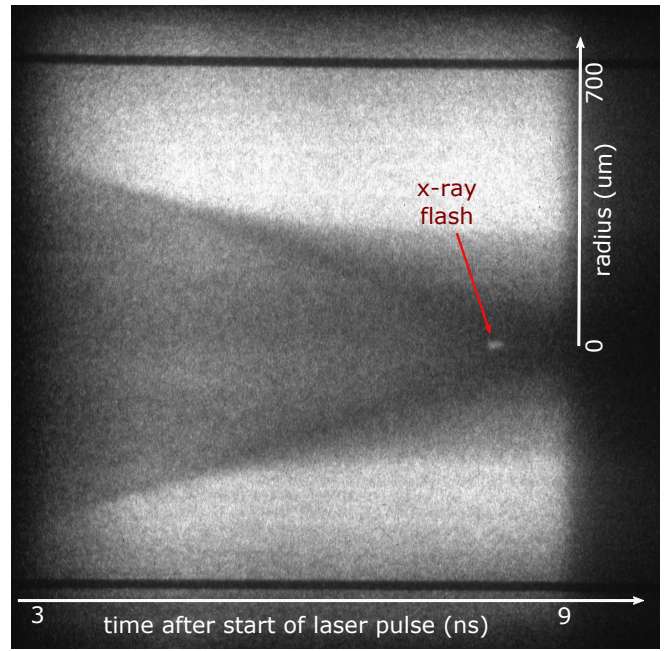}\end{center}
\caption{X-ray streak radiograph, NIF shot N140529-2 (high drive).}
\label{fig:rg140529}
\end{figure}

\section{Determination of absolute Hugoniot and opacity data from radiography}
As discussed previously \cite{Swift2018},
the reconstructed radius-time distribution of mass density $\rho(r,t)$ gives 
an absolute measurement of the shock Hugoniot over a range of pressures,
from the position of the shock $r_s(t)$ and hence its speed $u_s(t)$,
and the mass density immediately behind the shock,
$\rho_s(t)=\rho(r_s(t),t)$.
Simultaneous knowledge of $u_s$ and $\rho_s$ gives the complete mechanical
state behind the shock by solving the Rankine-Hugoniot relations \cite{shock}
representing the conservation of mass, momentum, and energy across the shock.
The state ahead of the shock is known, leaving five quantities to be
determined: $\rho$, pressure $p$, internal energy $e$, 
particle speed $u_p$, and shock speed $u_s$.
If any two of these quantities are measured, the Rankine-Hugoniot equations
determine the rest.
In particular,
$
p = p_0 + \left(v_0-v\right)u_s^2/v_0^2
$
where $v=1/\rho$ and subscript `0' denotes material ahead of the shock.
The Hugoniot state on the Hugoniot can thus be deduced directly from
the mass density distribution without reference to a standard material:
an absolute measurement.

Given distributions of mass density $\rho(\vec r,t)$ and 
opacity $\sigma(\vec r,t)$ in the object, 
the signal along any path from the source $\vec r_s$ to the detector $\vec r_d$
at any instant of time is given by the integral of attenuation $\mu=\rho\sigma$
through the object.
Given some radiographically-visible feature near the outside of the bead,
defining the enclosed mass,
the time-variation can be analyzed to 
determine changes in opacity independently from changes in density
\cite{Swift2021}.
As before, with a single marker layer, we had to use a model to account for the
isentropic variation of opacity with compression behind the shock.

We previously analyzed the streak radiographs by optimizing parameters
in models of the distribution of mass density $\rho(r,t)$, obtaining
a tabulated variation of opacity at the shock front $\sigma_s(t)$.
The density model was expressed in terms of the variation between the
trajectories of the shock and the marker, defined as analytic functions
with parameters included in the optimization.
The radiographs also include data on the EOS in the isentropic flow region
behind the shock.
In investigating how we might extract this off-Hugoniot data,
we found that we could parameterize the radiographic model more efficiently
-- with fewer parameters, and leading to a lower uncertainty -- by expressing
the mass density along the shock front in terms of the shock speed $u_s$ 
instead of time. $u_s$ is simply the derivative of the shock trajectory.
The radial derivative of density at the shock front can be related to
the acceleration of the shock, $\dot u_s$, and the isentropic sound speed $c$,
by
\begin{equation}
\dot u_s = \pdiffl \rho r \left.\pdiffl p\rho\right|_s
   \left[c(u_s)+u_p(u_s)-u_s\right] \pdiffl{u_s}{p_s}
\label{eq:acc}
\end{equation}
where $\partial p/\partial\rho|_s=c^2$,
and so we used an optimizable function $c(u_s)$ to capture the radial density
gradient.
We refer below to constructing $\rho(r,t)$ indirectly via 
$\rho(u_s)$ and $c(u_s)$ as Hugoniot functions.
We also used an optimizable function to describe the opacity in the
shocked state;
informed by comparisons with theory that were made when investigating
the results for polystyrene \cite{Kritcher2020},
we chose a Fermi-like function of shock pressure,
the pressure being obtained from the Hugoniot relations 
using the shock speed and mass density.
Low-order polynomials were found adequate for the other fitting functions
over the range of the data.
More details and comparisons of alternative analysis methods are
described elsewhere \cite{Swift2022}.

Constructed in this way, the functions over which parameters are optimized
are properties of the sample material and do not depend on the specific
experiment, in contrast to parameterizing $\rho(r,t)$ explicitly.
It is then possible to obtain the solution that best fits multiple
experiments simultaneously.

\section{Analysis of experimental data for diamond}
The streak radiograph was analyzed to deduce
the radius-time distribution of mass density,
represented using smooth functions as described above.
The analysis was performed in several different ways to explore the sensitivity
\cite{Swift2022}.
The locus of the shock was represented well by the function
$
r_s(t)=\alpha(t_c-t)^\beta
$
with parameters $\alpha$, $\beta$, and $t_c$.

Given a set of fitting parameters, 
the Hugoniot was obtained directly from the function $\rho_s(u_s)$,
the sound speed along the Hugoniot similarly from $c(u_s)$,
and the opacity from the function $\sigma(p)$.
The goodness-of-fit of the simulated radiograph to the data was used to assign a probability
to the model.
By perturbing the fitting parameters about the best fit, 
properties were deduced with corresponding probability,
and loci were accumulated as probability amplitudes.
The nominal best fit for each property was taken to be 
the locus of maximum likelihood,
very similar to the corresponding from the best-fitting parameters.
Statistical fitting uncertainties were obtained as contours 
from the probability distribution.
Systematic uncertainties from the
uncertainty in instantaneous sweep rate of the streak camera and magnification
affect the absolute values of each locus, 
but perturbed the shape to a much smaller degree,
as found previously \cite{Swift2018,Swift2021}. 

\begin{figure}
\begin{center}\includegraphics[scale=0.70]{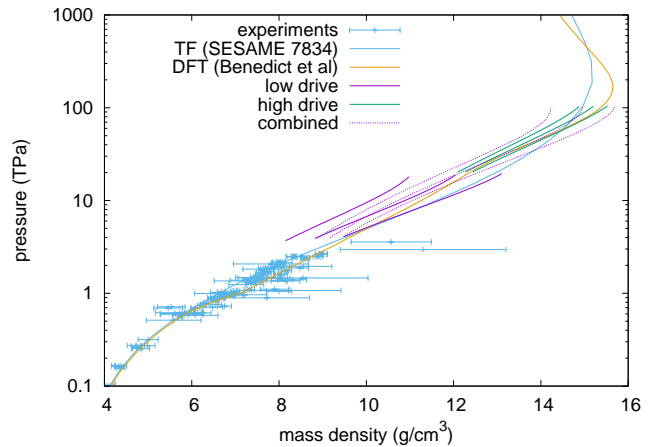}\end{center}
\caption{Principal Hugoniot for diamond mass density -- pressure,
   from the low and high drive shots separately and combined.
   Thin lines are $1\sigma$ uncertainty contours.
   Also shown are predictions from EOS based on a TF treatment of
   the electrons \cite{ses7834}
   and on density functional theory (DFT) calculations incorporating the
   effects of shell structure \cite{Benedict2014}.
   Points with uncertainty bars are previous experimental measurments
   \cite{Pavlovskii1971,Kondo1983,Bradley2004,Nagao2006,Brygoo2007,Hicks2008,McWilliams2010,Knudson2008,Gregor2017}.}
\label{fig:hugdp}
\end{figure}

\begin{figure}
\begin{center}\includegraphics[scale=0.70]{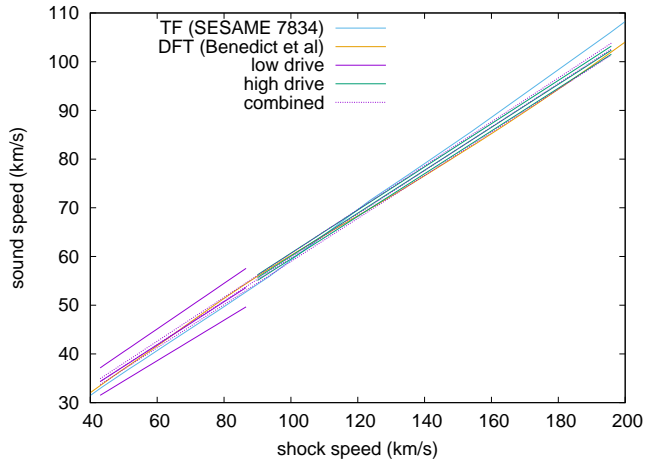}\end{center}
\caption{Sound speed along the principal Hugoniot for diamond,
   from the low and high drive shots separately and combined.
   Thin lines are $1\sigma$ uncertainty contours.
   Also shown are predictions from EOS based on a TF treatment of
   the electrons \cite{ses7834}
   and on density functional theory (DFT) calculations incorporating the
   effects of shell structure \cite{Benedict2014}.}
\label{fig:hugusc}
\end{figure}

\begin{figure}
\begin{center}\includegraphics[scale=0.70]{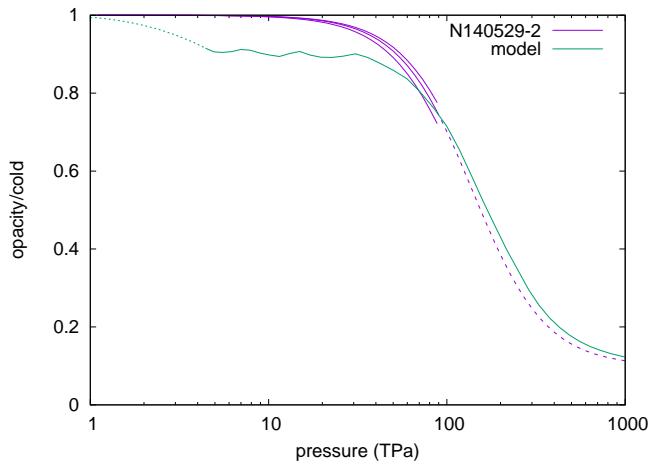}\end{center}
\caption{Opacity of carbon to 10.2\,keV x-rays
   along the principal shock Hugoniot for diamond.
   Thin lines are $1\sigma$ uncertainty contours.
   The dashed region is the extrapolation of the nominal fit to higher pressures.
   Also shown are predictions from the {\sc atomic} model \cite{atomic},
   with states calculated from the same EOS \cite{Benedict2014}.
   The dashed region of the model is interpolation between the ambient opacity
   and the first calculated value at 10\,eV, as discussed in the text.}
\label{fig:opac}
\end{figure}

We were able to obtain data along the principal Hugoniot between 3 and 80\,TPa.
The statistical uncertainty was substantially higher for the low-drive shot,
consistent with the lower signal level in the radiograph.
Fitting both shots together gave a significant reduction in the uncertainty
at low pressures.
The shape was in significantly better agreement with the theoretical EOS
constructed to account for shell effects \cite{Benedict2014}
than one based on TF predictions \cite{ses7834}
(Fig.~\ref{fig:hugdp}).
For the variation of sound speed with shock speed,
we were not able to detect any significant variation from a straight line,
consistent with both EOS models, though at high pressures the data
favored the shell structure EOS
No significant variation in opacity was found in the low-drive shot.
In the high-drive shot, the opacity was deduced to drop by $\sim$20\%\ between
10 and 80\,TPa.
The most rigorous prediction available to us was the {\sc atomic} model
\cite{atomic}, which accounts for detailed configurations of excited electrons.
Plasma opacity theory has difficulty at low temperatures, 
and values below 10\,eV are interpolations between the ambient opacity
and the first calculated value.
Combined with the shock Hugoniot states from the shell structure EOS,
the opacity was predicted to drop consistently with the variation deduced
from the experiment.
(Figs~\ref{fig:hugusc} and \ref{fig:opac}.)

\section{Conclusions}
These experiments have again pushed the limits of pressures accessible
in laboratory measurements of the equation of state, now approaching 100\,TPa.
Compared with our previously-reported experiments on polystyrene,
these measurements on higher-density diamond were affected less by the
drop in x-ray opacity induced by shock heating, even though the peak
pressure was over a factor of two higher and the peak density 3.5 times higher,
because of the greater starting density and stiffness of diamond.
Together with improvements in the method of analysis, the resulting data
were more consistent with theoretical predictions accounting for
electronic shell structure, rather than TF theory.
As before, the measurements are absolute rather than with respect to
a standard material of assumed equation of state, 
in contrast to previous measurements in this regime which employed
nuclear detonations as an energy source \cite{Ragan1980},
and were more susceptible to preheating from radiation in a planar
configuration.
The modified method of analysis also made it possible to deduce the
sound speed along the Hugoniot, providing a direct connection to
the Gr\"uneisen parameter and off-Hugoniot states.
Together with the previous measurements on polystyrene
\cite{Doeppner2018,Kritcher2020}, these data directly probe states
occurring in the envelope of white dwarf stars,
and show the importance of electronic shell structure effects.

\section*{Acknowledgments}
This work was performed
under the auspices of
the U.S. Department of Energy under contract DE-AC52-07NA27344.
R.W.F. acknowledges support from the Department of Energy, National Nuclear Security Administration Award DE-NA0003842, and the Department of Energy, Office of Science, Office of Fusion Energy Sciences Award DE-SC0018298.

The data that support the findings of this study are available from the corresponding author upon reasonable request.


\begin{thebibliography}{10}
\bibitem{AllendePrieto2002}{C.~Allende Pietro, D.L.~Lambert, and M.~Asplund, Astrophys. J. {\bf 573}, L137 (2002).}
\bibitem{vanHorn1968}{H.M.~van Horn, Astroph. J., {\bf 151}, 227 (1968).}
\bibitem{Horowitz2007}{C.J.~Horowitz, D.K.~Berry, and E.F.~Brown
   Phys. Rev. E {\bf 75}, 066101 (2007).}
\bibitem{Fontaine2001}{G.~Fontaine, P.~Brassard, and P.~Bergeron,
   Pub. Astron. Soc. Pacific {\bf 113}, 782, 409–435 (2001).}
\bibitem{shock}{For example,
   R.G.~McQueen et al, 
   in R.~Kinslow (Ed.),
   ``High Velocity Impact Phenomena''
   (Academic Press, New York, 1970).}
\bibitem{dac}{For example, A.D.~Chijioke,
   W.J.~Nellis, A.~Soldatov, and I.F.~Silvera,
   {\it The ruby pressure standard to 150 GPa},
   J.~Appl. Phys. {\bf 98}, 114905 (2005).}
\bibitem{Kritcher2014}{A.L.~Kritcher et al, 
   High Energy Density Phys. {\bf 10}, pp~27--34 (2014).}
\bibitem{Doeppner2018}{T.~D\"oppner, D.C.~Swift, A.L.~Kritcher, B.~Bachmann,
   G.W.~Collins, D.A.~Chapman, J.~Hawreliak, D.~Kraus, J.~Nilsen, S.~Rothman,
   L.X.~Benedict, E.~Dewald, D.E.~Fratanduono, J.A.~Gaffney, S.H.~Glenzer,
   S.~Hamel, O.L.~Landen, H.J.~Lee, S.~LePape, T.~Ma, M.J.~MacDonald,
   A.G.~MacPhee, D.~Milathianaki, M.~Millot, P.~Neumayer, P.A.~Sterne,
   R.~Tommasini, and R.W.~Falcone,
   Phys. Rev. Lett. {\bf 121}, 025001 (2018).}
\bibitem{Swift2018}{D.C.~Swift, A.L.~Kritcher, J.A.~Hawreliak, A.~Lazicki,
   A.~MacPhee, B.~Bachmann, T.~D\"oppner, J.~Nilsen, G.W.~Collins, S.~Glenzer,
   S.D.~Rothman, D.~Kraus, and R.W.~Falcone,
   Rev. Sci. Instrum. {\bf 89}, 053505 (2018).}
\bibitem{Kritcher2020}{A.L.~Kritcher, D.C.~Swift, T.~D\"oppner, B.~Bachmann, L.X.~Benedict, G.W.~Collins, J.L.~DuBois, F.~Elsner, G.~Fontaine, J.A.~Gaffney, S.~Hamel, A.~Lazicki, W.R.~Johnson, N.~Kostinski, D.~Kraus, M.J.~MacDonald, B.~Maddox, M.E.~Martin, P.~Neumayer, A.~Nikroo, J.~Nilsen, B.A.~Remington, D.~Saumon, P.A.~Sterne, W.~Sweet, A.A.~Correa, H.D.~Whitley, R.W.~Falcone, and S.H.~Glenzer,
   Nature {\bf 584}, 7819, pp~51-54 (2020).}
\bibitem{Swift2021}{D.C.~Swift et al, Rev. Sci. Instrum. {\bf 92}, 063514 (2021).}
\bibitem{hohlraum}{S.W.~Haan et al, 
   Nucl.~Fusion {\bf 44}, S171 (2004).}
\bibitem{icf}{J.~Lindl,
   Phys. Plasmas 2, 3933 (1995).}
\bibitem{Hopkins2015}{L.F.~Berzak Hopkins et al, 
   Phys. Rev. Lett. {\bf 114}, 175001 (2015).}
\bibitem{Dewald2004}{E.L.~Dewald et al, 
   Rev. Sci. Instrum. {\bf 75}, 3759 (2004).}
\bibitem{Swift2022}{D.C.~Swift, A.L.~Kritcher, A.L.~Lazicki, N.~Kostinski, B.R.~Maddox, M.E.~Martin, T.~D\"oppner, J.~Nilsen, and H.D.~Whitley,
   {\tt arXiv:2203.08891} (2022).}
\bibitem{xbt}{E.L.~Dewald et al, 
   Phys. Rev. Lett {\bf 111}, 235001 (2013).}
\bibitem{hydra}{M.M.~Marinak et al, 
   Phys. Plasmas {\bf 5}, 1125 (1998).}
\bibitem{ses7592}{J.~Barnes and S.~Lyon (Los Alamos National Laboratory),
 documentation for {\sc sesame} EOS 7592 (1988).}
\bibitem{opal}{C.A.~Iglesias, Astrophys. J. {\bf 464}, 943 (1996).}
\bibitem{atomic}{P.~Hakela, M.E.~Sherrill, S.~Mazevet, J.~Abdallah Jr., J.~Colgan, D.P.~Kilcrease, N.H.~Magee, C.J.~Fontes, and H.L.~Zhang,
   J.~Quantitative Spectroscopy and Radiative Transfer {\bf 99}, 265-271 (2006).}
\bibitem{Marsh1980}{S.P.~Marsh (Ed), {\it LASL Shock Hugoniot Data}
   (University of California, Berkeley, 1980).}
\bibitem{tf}{L.H.~Thomas,
   Proc. Cambridge Phil. Soc. {\bf 23}, 5, 542–548 (1927);
   E.~Fermi,
   Rend. Accad. Naz. Lincei. {\bf 6}, 602–607 (1927).}
\bibitem{Crockett2006}{S.~Crockett, documentation for {\sc sesame} EOS 7834,
   Los Alamos National Laboratory (2006).}
\bibitem{Liberman1979}{D.A.~Liberman, Phys. Rev.~B {\bf 20}, 12, 4981 (1979).}
\bibitem{Benedict2014}{L.X.~Benedict, K.P.~Driver, S.~Hamel, B.~Militzer, T.~Qi, A.A.~Correa, A.~Saul, and E.~Schwegler,
   Phys. Rev. B {\bf 89}, 224109 (2014).}
\bibitem{ses7834}{S.~Crockett (Los Alamos National Laboratory),
   documentation for {\sc sesame} 7834 (2006).}
\bibitem{Wilson2006}{B.~Wilson, V.~Sonnad, P.~Sterne, and W.~Isaacs, J.~Quant. Spectrosc. Radiat. Transfer {\bf 99}, 658 (2006).}
\bibitem{Pavlovskii1971}{M.N.~Pavlovskii, Sov. Phys. Solid State {\bf 13}, 741 (1971).}
\bibitem{Kondo1983}{K.~Kondo and T.J.~Ahrens, Geophys. Res. Lett. {\bf 10}, 281 (1983).}
\bibitem{Bradley2004}{D.K.~Bradley et al, Phys. Rev. Lett. {\bf 93}, 19, 195506 (2004).}
\bibitem{Nagao2006}{H.~Nagao et al., Phys. Plasmas {\bf 13}, 052705 (2006).}
\bibitem{Brygoo2007}{S.~Brygoo et al., Nat. Mater. {\bf 6}, 274 (2007).}
\bibitem{Hicks2008}{D.G.~Hicks et al, Phys. Rev. B {\bf 78}, 174102 (2008).}
\bibitem{McWilliams2010}{R.S.~McWilliams et al, Phys. Rev. B {\bf 81}, 014111 (2010).}
\bibitem{Knudson2008}{M.~Knudson et al, Science {\bf 322}, 1822 (2008).}
\bibitem{Gregor2017}{M.C.~Gregor et al, Phys. Rev. B {\bf 95}, 144114 (2017).}
\bibitem{Ragan1980}{For example, C.E.~Ragan III, Phys. Rev. A {\bf 21}, 458 (1980).}
\end{thebibliography}
\end{document}